\documentstyle[12pt]{article}

\newcommand {\e} {\mbox{\rm e}}

\newcommand {\nn}    {\nonumber}
\newcommand {\vs}[1]  { \vspace*{#1 cm} }

\newcounter{eq}
\newcounter{sc}

\newcommand {\MPL}  {Mod. Phys. Lett.}
\newcommand {\NP}   {Nucl. Phys.}
\newcommand {\PL}   {Phys. Lett.}
\newcommand {\PR}   {Phys. Rev.}

\newcommand {\JHEP}  {JHEP}

\newcommand {\ZP}  {Z. Phys.}

\def\overleftrightarrow#1{\vbox{\ialign{##\crcr
 $\leftrightarrow$\crcr\noalign{\kern-1pt\nointerlineskip}
 $\hfil\displaystyle{#1}\hfil$\crcr}}}

\newlength{\minitwocolumn}
\begin{document}

\begin{flushright}
EDO-EP-45\\
February, 2003\\
hep-th/0302203\\
\end{flushright}
\vspace{30pt}

\pagestyle{empty}
\baselineskip15pt

\begin{center}
{\large\bf Covariant Matrix Model of Superparticle

in the Pure Spinor Formalism
 \vskip 1mm
}

\vspace{20mm}

Ichiro Oda
          \footnote{
          E-mail address:\ ioda@edogawa-u.ac.jp
                  }
\\
\vspace{10mm}
          Edogawa University,
          474 Komaki, Nagareyama City, Chiba 270-0198, JAPAN \\

\end{center}


\vspace{15mm}
\begin{abstract}

On the basis of the Berkovits pure spinor formalism of
covariant quantization of supermembrane, we attempt to
construct a M(atrix) theory which is covariant under $SO(1,10)$
Lorentz group. We first construct a bosonic M(atrix) theory
by starting with the first-order formalism of bosonic
membrane, which precisely gives us a bosonic sector of 
M(atrix) theory by BFSS. Next we generalize this method to the 
construction of M(atrix) theory of supermembranes.  
However, it seems to be difficult to obtain a covariant and 
supersymmetric M(atrix) theory from the Berkovits pure spinor 
formalism of supermembrane because of the matrix character of 
the BRST symmetry. Instead, in this paper, we construct a 
supersymmetric and covariant matrix model of 11D superparticle,
which corresponds to a particle limit of covariant M(atrix) theory.
By an explicit calculation, we show that the one-loop effective
potential is trivial, thereby implying that this matrix model
is a free theory at least at the one-loop level.

\vspace{15mm}

\end{abstract}

\newpage
\pagestyle{plain}
\pagenumbering{arabic}


\rm

\section{Introduction}

Since the advent of M(atrix) theory by BFSS  \cite{BFSS}, there has
been a strong desire to construct a manifestly Lorentz covariant
M(atrix) theory, but no one has succeeded in constructing such a
theory thus far.

Although M(atrix) theory has been derived from the low-energy effective
action of D-particles which is obtained via the dimensional reduction
from the maximally supersymmetric Yang-Mills theory in ten dimensions,
this theory can be also interpreted as a regularized supermembrane
theory in the light-cone gauge \cite{de Witt}. Then, it is natural to
start with the supermembrane action in eleven dimensions and quantize
it in a covariant manner in order to obtain the covariant M(atrix)
theory. However, as is well known, the fermionic kappa symmetry,
which is used to reduce the number of fermionic degrees of freedom
by half, has given us a difficulty in covariant quantization of
the supermembrane action. 
 
Recently, there has been an interesting progress by Berkovits in the
covariant quantization of the Green-Schwarz superstrings \cite{GS}
using the pure spinors \cite{Ber1, Ber2, Ber3, Ber4, Ber5}. 
One of the key ingredients in the Berkovits approach is the existence 
of the BRST charge $Q_{BRST} = \oint \lambda^\alpha d_\alpha$ 
where $\lambda^\alpha$ are pure spinors satisfying the pure spinor 
equations $\lambda^\alpha \Gamma^m_{\alpha \beta} \lambda^\beta = 0$ and 
$d_\alpha \approx 0$ are the fermionic constraints associated with 
the kappa symmetry. It is remarkable that this approach provides us 
the same cohomology as the BRST charge of the Neveu-Schwarz-Ramond 
formalism \cite{Neveu} and the correct tree amplitudes of superstrings 
with keeping the Lorentz covariance of the theory. Afterwards, the
Berkovits approach has been investigated from various different
viewpoints \cite{Oda1, Ber6, Ber7, Trivedi, Oda2, Grassi, Kazama}.
In particular, more recently, the generalization of this approach 
to supermembrane has be done by Berkovits \cite{Ber8}.

Combining the above-mentioned two observations, we are naturally led to 
think that we could make use of the Berkovits pure spinor formalism 
to construct a covariant
M(atrix) theory since the covariant quantization of supermembrane
has been made and the difficulty of the quantization associated with
the kappa symmetry has been resolved in the pure spinor formalism. 
Actually, Berkovits has proposed such an interesting idea in the 
conference of Strings 2002 \cite{Ber9}, but it is a pity that this 
work has not been completed so far as long as I know.

In this paper, we pursue this idea and attempt to construct
a covariant M(atrix) theory by using the pure spinor formalism of
supermembrane \cite{Ber8}. However, we will see that the 
construction of a covariant M(atrix) theory is rather difficult 
owing to the existence of the BRST invariance $Q_{BRST}$ which is now
promoted to a matrix symmetry (like a local gauge symmetry) in the 
matrix model. In this paper, we will explain in detail why it is 
difficult to apply the Berkovits formalism to the construction of the 
covariant M(atrix) model.

Thus, instead of constructing a covariant M(atrix) theory,
we present how to construct a covariant matrix model of 
superparticle in eleven dimensions \cite{Brink} which in some sense
corresponds to a particle limit of a covariant M(atrix) theory.

This paper is organized as follows. In section 2, as a warmup,
we construct a bosonic M(atrix) theory by starting with the first-order 
formalism of bosonic membrane. In section 3, we generalize the method
to supermembrane and attempt to construct a covariant M(atrix)
theory from the pure spinor formalism of supermembrane by Berkovits. 
Here we find a difficulty of constructing M(atrix) theory which is
invariant under the BRST symmetry. Hence, instead we turn to the 
construction of a covariant matrix model of superparticle invariant 
under both the supersymmetry and the BRST symmetry. Furthermore, 
in section 4, we calculate the one loop effective potential and 
show that our matrix model is a free theory owing to the lack of
the potential term. The final section is devoted to the conclusion.

\rm
\section{Bosonic M(atrix) theory}

In this section, we shall construct a bosonic M(atrix) theory since
this construction gives us a good exercise in attempting to construct
a M(atrix) theory of supermembrane based on the pure spinor
formalism. In addition, we can clearly understand the difference
of the construction of a matrix model between the bosonic theory
and the supersymmetric one. 
A similar analysis has been thus far done from various different
contexts \cite{Bergshoeff, Santos, Fujikawa, Ishibashi, Yoneya}.

We begin with the well-known Nambu-Goto action of the bosonic
membrane in eleven dimensions in a flat space-time:
\begin{eqnarray}
S_{NG} = - T \int d^3 \sigma \sqrt{-g},
\label{1}
\end{eqnarray}
where $T$ is the membrane tension with dimension $(mass)^3$. 
And the induced metric and its determinant are respectively given by 
$g_{ij}= \partial_i x^a \partial_j x^b \eta_{ab}$, and $g = \det g_{ij}$. 
We take the Minkowskian metric signature $(-,+,+, \cdots,+)$. Moreover, 
the indices indicate $i, j = 0, 1, 2$ and $a, b, c = 1, 2, \cdots, 11$. 
We follow the notations and conventions of the Berkovits' paper \cite{Ber8}.

Let us perform the canonical quantization of the action (\ref{1}).
The canonical conjugate momenta of $x^a$ are derived as
\begin{eqnarray}
P_a &=& \frac{\partial S_{NG}}{\partial \dot{x}^a} \nn\\
&=& - T \sqrt{-g} g^{0j} \partial_j x_a \nn\\
&=& - T \sqrt{-g} (g^{00} \partial_0 x_a + g^{0I} \partial_I x_a),
\label{2}
\end{eqnarray}
where we have defined as $\dot{x}^a = \partial_0 x^a$ and $I, J = 1, 2$. 
{}From this expression, we have the primary constraints which generate
the world-volume reparametrization invariance as follows:
\begin{eqnarray}
{\cal{H}}_0 &=& \frac{1}{2T} P_a P^a + \frac{T}{2} h \approx 0, \nn\\
{\cal{H}}_I &=& P_a \partial_I x^a \approx 0,
\label{3}
\end{eqnarray}
where $h_{IJ} = \partial_I x^a \partial_J x^b \eta_{ab}$ and
$h = \det h_{IJ}$. Given the Poisson brackets
\begin{eqnarray}
\{P_a(\sigma^0, \vec{\sigma}), x^b (\sigma^0, \vec{\sigma}')\} 
= - \delta_a^b \delta^2(\vec{\sigma} - \vec{\sigma}'),
\label{4}
\end{eqnarray}
it is easy to show that the constraints constitute of the first-class
constraints as required:
\begin{eqnarray}
\{ {\cal{H}}_0 (\sigma^0, \vec{\sigma}), {\cal{H}}_0 (\sigma^0, \vec{\sigma}') \}
&=&  \Bigl[ {\cal{H}}_I(\vec{\sigma}) h(\vec{\sigma}) h^{IJ}(\vec{\sigma})
+ {\cal{H}}_I(\vec{\sigma}') h(\vec{\sigma}') h^{IJ}(\vec{\sigma}') \Bigr]
\partial_J \delta(\vec{\sigma} - \vec{\sigma}'), \nn\\
\{ {\cal{H}}_0 (\sigma^0, \vec{\sigma}), {\cal{H}}_I (\sigma^0, \vec{\sigma}') \}
&=&  \Bigl[ {\cal{H}}_0(\vec{\sigma}) + {\cal{H}}_0(\vec{\sigma}') \Bigr]
\partial_I \delta(\vec{\sigma} - \vec{\sigma}'), \nn\\
\{ {\cal{H}}_I (\sigma^0, \vec{\sigma}), {\cal{H}}_J (\sigma^0, \vec{\sigma}') \}
&=& {\cal{H}}_J(\vec{\sigma}) \partial_I \delta(\vec{\sigma} 
- \vec{\sigma}')  + {\cal{H}}_I(\vec{\sigma}') \partial_J \delta(\vec{\sigma} 
- \vec{\sigma}').
\label{5}
\end{eqnarray}

Since the Hamiltonian vanishes weakly, we can introduce the
extended Hamiltonian which is purely proportional to the constraints
\begin{eqnarray}
H &=& \int d^2 \vec{\sigma} \Bigl[ e^0 {\cal{H}}_0 + e^I {\cal{H}}_I 
\Bigr] \nn\\
&=& \int d^2 \vec{\sigma} \Bigl[ e^0 (\frac{1}{2T} P_a P^a + \frac{T}{2} h)
 + e^I P_a \partial_I x^a \Bigr],
\label{6}
\end{eqnarray}
where $e^0$ and $e^I$ are the Lagrange multiplier fields.
Via the Legendre transformation, we can obtain the first-order action:
\begin{eqnarray}
S_0 &=& \int d^3 \sigma P_a \partial_0 x^a - \int d \sigma^0 H \nn\\
&=& \int d^3 \sigma \Bigl[ P_a \partial_0 x^a - e^0 \bigl(\frac{1}{2T} 
P_a P^a + \frac{T}{2} h \bigr) - e^I P_a \partial_I x^a \Bigr].
\label{7}
\end{eqnarray}
Note that this action is very similar to the bosonic part of the
Berkovits action of supermembrane \cite{Ber8} in that both the actions are
in the first-order Hamiltonian form and invariant under only the
world-volume reparametrizations as local symmetries, 
so it is worthwhile to construct a bosonic matrix model from this
action. 
Actually, we will see that the construction of M(atrix) theory follows a
very similar path to the present bosonic formalism.

In order to construct a matrix model, we first perform the integration
over $P_a$ whose result is given by
\begin{eqnarray}
S_0 &=& \frac{T}{2} \int d^3 \sigma \Bigl[ \frac{1}{e^0} 
\bigl( \partial_0 x^a - e^I \partial_I x^a \bigr)^2 - e^0 h \Bigr] \nn\\
&=& \frac{T}{2} \int d^3 \sigma \Bigl[ \frac{1}{e^0} 
\bigl( \partial_0 x^a - e^I \partial_I x^a \bigr)^2 - \frac{1}{2} e^0 
\{ x^a, x^b \}^2 \Bigr],
\label{8}
\end{eqnarray}
where in the second equation we have introduced the Lie bracket 
defined as
\begin{eqnarray}
\{ X, Y \} = \varepsilon^{IJ} \partial_I X \partial_J Y.
\label{9}
\end{eqnarray}

Here let us try to understand the geometrical meaning of
the Lagrange multiplier fields, which can be done by comparing
the above action with the Polyakov action (which is at least classically
equivalent to the Nambu-Goto action (\ref{1}))
\begin{eqnarray}
S_P = T \int d^3 \sigma \Bigl( - \frac{1}{2} \sqrt{-g} g^{ij}
\partial_i x^a \partial_j x^b \eta_{ab} + \frac{1}{2} \sqrt{-g} \Bigr).
\label{10}
\end{eqnarray}
Then we can express the metric tensor in terms of the Lagrange
multiplier fields 
\begin{eqnarray}
g_{ij} &=& \left(
\begin{array}{cc}
e^I e^J h_{IJ} - (e^0)^2 h & h_{JK} e^K \\
h_{IL} e^L & h_{IJ}
\end{array}
\right), \nn\\
g^{ij} &=& \left(
\begin{array}{cc}
- \frac{1}{(e^0)^2 h} & \frac{e^J}{(e^0)^2 h} \\
\frac{e^I}{(e^0)^2 h} & h^{IJ} - \frac{e^I e^J}{(e^0)^2 h}
\end{array}
\right).
\label{11}
\end{eqnarray}

Next we will fix the reparametrization invariance by two gauge
conditions \footnote{For comparison with the case of
supermembrane in the next section, we will not take the 
light-cone gauge explicitly in what follows. }. 
The first choice of the gauge conditions is given by
\begin{eqnarray}
e^0 = \frac{1}{\sqrt{h}}, \ e^I = 0,
\label{12}
\end{eqnarray}
or equivalently, from (\ref{11}),
\begin{eqnarray}
g_{ij} = \left(
\begin{array}{cc}
- 1 & 0 \\
0  & h_{IJ}
\end{array}
\right).
\label{13}
\end{eqnarray}
With the gauge conditions (\ref{12}), the action (\ref{8}) reduces to
\begin{eqnarray}
S_0 = \frac{T}{2} \int d \sigma^0 \int d^2 \sigma \sqrt{h}
\Bigl[ \bigl( \partial_0 x^a \bigr)^2 - \frac{1}{2 h} 
\{ x^a, x^b \}^2 \Bigr].
\label{14}
\end{eqnarray}

Finally, we make the following replacements
\begin{eqnarray}
\int d^2 \sigma \sqrt{h} &\rightarrow& Tr, \nn\\
\frac{1}{\sqrt{h}} \{ x^a, x^b \} &\rightarrow& i [ x^a, x^b ].
\label{15}
\end{eqnarray}
Consequently, we arrive at a matrix model of the bosonic membrane
\begin{eqnarray}
S_0 = \int d \tau Tr
\Bigl\{ \frac{1}{2} \bigl( \partial_\tau x^a \bigr)^2
+ \frac{1}{4}  [ x^a, x^b ]^2 \Bigr\},
\label{16}
\end{eqnarray}
where we have set $T = 1$ and $\sigma^0 = \tau$. This matrix model 
describes a matrix model of the bosonic membrane. 

We can also select another form of the gauge conditions $e^0 =
\frac{1}{\sqrt{h}}$ and $\ e^I = \frac{1}{\sqrt{h}} \varepsilon^{IJ}
\partial_J A_0$.
Then, the action (\ref{8}) takes the form
\begin{eqnarray}
S_0 = \frac{T}{2} \int d \sigma^0 \int d^2 \sigma \sqrt{h}
\Bigl[ \bigl( \partial_0 x^a + \frac{1}{\sqrt{h}} \{ A_0, x^a \}
\bigr)^2 
- \frac{1}{2 h} \{ x^a, x^b \}^2 \Bigr].
\label{17}
\end{eqnarray}
With the replacements (\ref{15}), we have a matrix model
\begin{eqnarray}
S_0 = \int d \tau Tr
\Bigl\{ \frac{1}{2} \bigl( D_\tau x^a \bigr)^2
+ \frac{1}{4}  [ x^a, x^b ]^2 \Bigr\},
\label{18}
\end{eqnarray}
where $D_\tau x^a = \partial_\tau x^a + i [A_\tau, x^a]$ and 
$A_\tau \equiv A_0$. 
This matrix model is obviously invariant under the $SU(N)$
gauge symmetry
\begin{eqnarray}
x^a &\rightarrow& x^{\prime a}  = U^{-1} x^a U, \nn\\
A_\tau &\rightarrow& A^{\prime}_\tau = U^{-1} A_\tau U - i U^{-1} 
\partial_\tau U.
\label{19}
\end{eqnarray}
With the gauge condition $A_\tau = 0$, this matrix model reduces
to the previous matrix model (\ref{16}).
Note that if we selected the light-cone gauge, the matrix model 
(\ref{18}) would become equivalent to the bosonic part of M(atrix) 
theory by BFSS except irrelevant dimensional factors and numerical 
constants \cite{BFSS} \footnote {M(atrix) theory manifestly depends on the 
background flat metric, so it is not a background independent formalism.
See \cite{Oda3} for the pioneering works of the background independent 
matrix models. }.
In this way, we can obtain the bosonic M(atrix) theory by starting
with the bosonic membrane action and utilizing the first-order
Hamiltonian formalism.

\rm
\section{A covariant matrix model of 11D superparticle}

We now turn our attention to an attempt of the construction of 
a covariant M(atrix) theory of supermembrane in the pure spinor formalism
and point out a difficulty of it.
Then we construct a new matrix model of superparticle in the pure 
spinor formalism.
 
Before doing so, let us start by reviewing the pure spinor
formalism of supermembrane \cite{Ber8}. From now on, we consider 
only the flat membrane such as toroidal membrane where the scalar 
density $\sqrt{h}$ can be set to unity.

The first-order Hamiltonian action of supermembrane reads
\begin{eqnarray}
S = \int d^3 \sigma \Bigl[ P_c \Pi_0^c + L_{WZ} 
+ e^0 \Bigl( P_c P^c + \det \bigl(\Pi_I^c \Pi_{J c} \bigr) \Bigr)
+ e^I P_c \Pi_I^c \Bigr],
\label{20}
\end{eqnarray}
where $\Pi_i^c = \partial_i x^c + \frac{i}{2} \theta \Gamma^c 
\partial_i \theta$ and $L_{WZ}$ denotes the Wess-Zumino term whose 
concrete expression takes the form
\begin{eqnarray}
L_{WZ} = \frac{i}{4} \varepsilon^{ijk} \theta \Gamma_{cd} \partial_i
\theta \Bigl( \Pi_j^c \Pi_k^d - \frac{i}{2} \Pi_j^c \theta \Gamma^d
\partial_k \theta - \frac{1}{12} \theta \Gamma^c \partial_j \theta 
\theta \Gamma^d \partial_k \theta  \Bigr),
\label{21}
\end{eqnarray}
where we define as $\varepsilon_{012}= - \varepsilon^{012}=
+1$ and $\varepsilon^{0IJ}= -\varepsilon^{IJ}$.
This action is invariant under the kappa symmetry and the global
space-time supersymmetry as well as the world-volume reparametrizations.
The primary constraints consisting of 16 first-class and 16 second-class 
constraints appear when we evaluate the canonical conjugate momenta
$p_\alpha$ of the spinor fields $\theta^\alpha$, which are given by
\begin{eqnarray}
d_\alpha &\equiv& p_\alpha - \frac{\partial^R S}{\partial 
\dot{\theta}^\alpha} \nn\\
&=& p_\alpha - \frac{i}{2} P^c (\Gamma_c \theta)_\alpha 
+ \frac{i}{4} \varepsilon^{IJ} (\Gamma_{cd} \theta)_\alpha
\Bigl( \Pi_I^c \Pi_J^d - \frac{i}{2} \Pi_I^c \theta \Gamma^d
\partial_J \theta - \frac{1}{12} \theta \Gamma^c \partial_I \theta 
\theta \Gamma^d \partial_J \theta \Bigr) \nn\\
&+& \frac{1}{8} \varepsilon^{IJ} \theta \Gamma_{cd} \partial_I \theta
\Bigl( \Pi_J^d - \frac{i}{6} \theta \Gamma^d \partial_J \theta \Bigr)
(\Gamma^c \theta)_\alpha  \nn\\
&\approx& 0,
\label{22}
\end{eqnarray}
where the superscript $R$ on $\frac{\partial^R S}{\partial 
\dot{\theta}^\alpha}$ denotes the right differentiation.
These constraints satisfy the following Poisson bracket
\begin{eqnarray}
\{ d_\alpha(\sigma^0, \vec{\sigma}), d_\beta(\sigma^0, \vec{\sigma}')\}
= \Bigl[-i P_c \Gamma_{\alpha\beta}^c + \frac{i}{2} \varepsilon^{IJ} 
\Pi_{I c} \Pi_{J d} \Gamma_{\alpha\beta}^{cd} \Bigr]
\delta^2(\vec{\sigma} - \vec{\sigma}').
\label{23}
\end{eqnarray}
In deriving this equation, we need to use the eleven dimensional 
Fierz identity 
$\Gamma_{(\alpha\beta}^b \Gamma_{\gamma\delta)}^{cd} \eta_{bc} = 0$
and the Poisson brackets
\begin{eqnarray}
\{\tilde{P}_c(\sigma^0, \vec{\sigma}), x^d (\sigma^0, \vec{\sigma}')\} 
&=& - \delta_c^d \delta^2(\vec{\sigma} - \vec{\sigma}'), \nn\\
\{ p_\alpha(\sigma^0, \vec{\sigma}), \theta^\beta(\sigma^0, \vec{\sigma}')\}
&=& \delta_\alpha^\beta \delta^2(\vec{\sigma} - \vec{\sigma}'),
\label{24}
\end{eqnarray}
where $\tilde{P}_c$, the conjugate momenta of $x^c$, are defined as
\begin{eqnarray}
\tilde{P}_c &\equiv& \frac{\partial S}{\partial \dot{x}^c} \nn\\
&=& P_c + \frac{i}{2} \varepsilon^{IJ} \theta \Gamma_{cd} \partial_I \theta
\Bigl( \Pi_J^d - \frac{i}{4} \theta \Gamma^d \partial_J \theta \Bigr).
\label{25}
\end{eqnarray}

Since we cannot quantize the action (\ref{20}) covariantly owing to
the kappa symmetry, Berkovits has proposed a pure spinor action, which is of form
\begin{eqnarray}
S &=& \int d^3 \sigma \Bigl[ P_c \Pi_0^c + L_{WZ} 
+ d_\alpha \partial_0 \theta^\alpha + w_\alpha \partial_0 \lambda^\alpha
- \frac{1}{2} \Bigl( P_c P^c + \det \bigl(\Pi_I^c \Pi_{J c} \bigr)\Bigr)\nn\\
&+& (d \Gamma_c \partial_I \theta) \Pi_J^c \varepsilon^{IJ}
+ (w \Gamma_c \partial_I \lambda) \Pi_J^c \varepsilon^{IJ}
- i \varepsilon^{IJ} (w \Gamma_c \partial_I \theta)(\lambda
\Gamma^c \partial_J \theta)  + i \varepsilon^{IJ}(w_\alpha \partial_I
\theta^\alpha)(\lambda_\beta \partial_J \theta^\beta) \nn\\
&+& e^I (P_c \Pi_I^c + d_\alpha \partial_I \theta^\alpha + w_\alpha 
\partial_I \lambda^\alpha) \Bigr],
\label{26}
\end{eqnarray}
where $d_\alpha$ is defined as in (\ref{22}).
In this action, the kappa symmetry has been already gauge-fixed
covariantly, whereas the shift symmetries of the world-volume
reparametrizations are still remained. (The lapse symmetry is gauge-fixed
to $e^0 = -\frac{1}{2}$.) This action is invariant under the BRST
transformation $Q_B = \int d^2 \sigma \lambda^\alpha d_\alpha$. As a
peculiar feature of supermembrane, additional constraints
\begin{eqnarray}
\lambda \Gamma^c \lambda = 0, \ (\lambda \Gamma^{cd} \lambda) \Pi_{Jc}
=0, \ \lambda_\alpha \partial_J \lambda^\alpha = 0
\label{27}
\end{eqnarray}
are required to guarantee the BRST invariance of the action 
and the nilpotence of the BRST transformation. Note that the 
constraints (\ref{27}) break the covariance on the world-volume
explicitly. 

Note that the bosonic part in the pure spinor action (\ref{26}) 
of supermembrane shares the same form as in the bosonic
membrane argued in the previous section, so as in the bosonic membrane, 
let us proceed to integrate over $P_c$ and choose the gauge conditions 
$e^I = - \varepsilon^{IJ} \partial_J A_0$ \footnote{Comparing (\ref{7})
and (\ref{26}), we notice that the definition of $e^0$ and $e^I$ in 
supermembrane differs from that in the bosonic membrane by the minus sign.}. 
As a result, the action (\ref{26}) reduces to the form
\begin{eqnarray}
S &=& \int d^3 \sigma \Bigl[ \frac{1}{2} (D_0 x^c + \frac{i}{2} 
\theta \Gamma^c D_0 \theta)^2 + L_{WZ} 
+ d_\alpha D_0 \theta^\alpha + w_\alpha D_0 \lambda^\alpha
- \frac{1}{2} \det \bigl(\Pi_I^c \Pi_{J c} \bigr) \nn\\
&+& (d \Gamma_c \partial_I \theta) \Pi_J^c \varepsilon^{IJ}
+ (w \Gamma_c \partial_I \lambda) \Pi_J^c \varepsilon^{IJ}
- i \varepsilon^{IJ} (w \Gamma_c \partial_I \theta)(\lambda
\Gamma^c \partial_J \theta) \nn\\
&+& i \varepsilon^{IJ}(w_\alpha \partial_I
\theta^\alpha)(\lambda_\beta \partial_J \theta^\beta) 
\Bigr],
\label{28}
\end{eqnarray}
where we have defined as $D_0 = \partial_0 - \varepsilon^{IJ}
\partial_J A_0 \partial_I$.

Via the replacements  (\ref{15}) (recall that we have set $h =1$) 
from the continuum theory to the matrix model, we obtain a covariant 
matrix model corresponding to the pure spinor action of supermembrane:
\begin{eqnarray}
S &=& \int d \tau Tr  \Bigg\{ \frac{1}{2} \left(D_\tau x^c + \frac{i}{4} 
(\theta \Gamma^c D_\tau \theta - D_\tau \theta \Gamma^c \theta) \right)^2 
+ d_\alpha D_\tau \theta^\alpha + w_\alpha D_\tau \lambda^\alpha \nn\\
&-& (\Gamma_c d)_\alpha \left( i [x^c, \theta^\alpha] 
- \frac{1}{2}(\Gamma^c \theta)_\beta \{\theta^\alpha, \theta^\beta \}
\right) 
- (\Gamma_c w)_\alpha \left( i [x^c, \lambda^\alpha] 
+ \frac{1}{2} (\Gamma^c \theta)_\beta [\lambda^\alpha, \theta^\beta]
\right) \nn\\
&-& \left(w_\alpha \lambda_\beta - (\Gamma_c w)_\alpha (\Gamma^c
 \lambda)_\beta \right)
\{\theta^\alpha, \theta^\beta \}  + L_{WZ} + L_{\det \pi^2}
\Bigg\},
\label{29}
\end{eqnarray}
where we have defined the covariant derivative 
as $D_\tau x^c = \partial_\tau x^c + i \left[A_\tau, x^c \right]$
as before, and
the curly bracket $\{\ , \ \}$ denotes the anti-commutator whereas
the square bracket $[\ , \ ]$ denotes the commutator. The last two terms
$L_{WZ}$ and $L_{\det \pi^2}$ come from the Wess-Zumino term and
$- \frac{1}{2} \det \bigl(\Pi_I^c \Pi_{J c} \bigr)$, respectively,
and involve the complicated expression.
Note that in moving the continuum theory to the matrix theory
we must pay attention to how to order various terms (in particular, in
$L_{WZ}$ and $L_{\det \pi^2}$). Our guiding principle is to order
the terms in order to keep symmetries of the theory as much as possible.

At this stage, compared with the bosonic membrane in the previous
section, we further have to impose the requirements of the supersymmetry and
the BRST invariance on the matrix model (\ref{29}). First, let us
consider the supersymmetry. This symmetry is a global symmetry, so
the matrix extension can be given by
\begin{eqnarray}
\delta x^c = \frac{i}{2} \theta \Gamma^c \epsilon, \ 
\delta \theta^\alpha = \epsilon^\alpha,
\label{30}
\end{eqnarray}
where the parameter $\epsilon$ is not a matrix but a mere number.
We have checked that under this supersymmetry the matrix model
(\ref{29}) is invariant except the Wess-Zumino term $L_{WZ}$.
To do so, we need to define $L_{\det \pi^2}$ in an appropriately
ordered form:
\begin{eqnarray}
L_{\det \pi^2} &=& \frac{1}{4} [x^a, x^b]^2 - \frac{i}{2}[x^a, x^b]
(\Gamma_a \theta)_\alpha [x_b, \theta^\alpha] \nn\\
&+& \frac{1}{8}[x^a, x^b](\Gamma_a \theta)_\alpha (\Gamma_b
 \theta)_\beta \{\theta^\alpha, \theta^\beta\}
- \frac{1}{8}(\Gamma_a \theta)_\alpha [x^b, \theta^\alpha]
(\Gamma^a \theta)_\beta [x_b, \theta^\beta] \nn\\
&+& \frac{1}{8}(\Gamma_a \theta)_\alpha [x^b, \theta^\alpha]
(\Gamma_b \theta)_\beta [x^a, \theta^\beta] 
+ \frac{i}{8}(\Gamma^a \theta)_\alpha [x^b, \theta^\alpha]
(\Gamma_{[b} \theta)_\beta (\Gamma_{a]} \theta)_\rho \{\theta^\beta,
\theta^\rho \} \nn\\
&+& \frac{1}{64} (\Gamma_{[a} \theta)_\alpha (\Gamma_{b]} \theta)_\beta
\{\theta^\alpha, \theta^\beta \} (\Gamma^a \theta)_\rho (\Gamma^b
\theta)_\sigma \{\theta^\rho, \theta^\sigma\},
\label{31}
\end{eqnarray}
where we have used the notation like $(\Gamma_{[a} \theta)_\alpha
(\Gamma_{b]} \theta)_\beta \equiv \frac{1}{2} \left( (\Gamma_a
\theta)_\alpha (\Gamma_b \theta)_\beta - (\Gamma_b \theta)_\alpha
(\Gamma_a \theta)_\beta \right)$. The reason why the Wess-Zumino term
is not invariant under the supersymmetry might be related to the fact
that this term breaks the $SU(N)$ gauge symmetry since it includes
not the covariant derivative $D_\tau$ but the ordinary derivative
$\partial_\tau$.  
Since the supersymmetry is an essential ingredient of our formalism,
we stick to keep this symmetry and drop the Wess-Zumino term $L_{WZ}$ 
from the matrix theory.

Furthermore, in case of supermembrane in the pure spinor formalism,
we must respect the BRST symmetry. The Berkovits'
BRST symmetry is simply given by $Q_B = \int d^2 \sigma \lambda^\alpha 
d_\alpha$.
Thus, using Eq. (\ref{15}) the matrix extension must take the form 
of $Q_B = Tr \lambda^\alpha d_\alpha$. Since we have dropped the 
Wess-Zumino term, $d_\alpha$ is now simply given by 
\begin{eqnarray}
d_\alpha = p_\alpha - \frac{i}{4} \left( P_c (\Gamma^c \theta)_\alpha
+ (\Gamma^c \theta)_\alpha P_c \right).
\label{32}
\end{eqnarray}
Then we have the following BRST transformation
\begin{eqnarray}
Q_B \theta^\alpha &=&  \lambda^\alpha, \nn\\
Q_B x^c &=& \frac{i}{4} \left( \theta \Gamma^c \lambda + \lambda
			 \Gamma^c \theta \right), \nn\\
Q_B d_\alpha &=& - \frac{i}{2} \left( P_c (\Gamma^c \lambda)_\alpha 
+ (\Gamma^c \lambda)_\alpha P_c  \right), \nn\\
Q_B w_\alpha &=&  d_\alpha.
\label{33}
\end{eqnarray}

Although this BRST symmetry has a rather simple form (essentially
is of the same form as in superparticle), this symmetry constrains 
the form of the action severely since all the fields are now promoted to
matrices.  Actually we can check that the BRST-invariant matrix model 
must take the form 
\begin{eqnarray}
S &=& \int d \tau Tr  \Bigg\{ \frac{1}{2} \left(D_\tau x^c + \frac{i}{4} 
(\theta \Gamma^c D_\tau \theta - D_\tau \theta \Gamma^c \theta) \right)^2 
+ d_\alpha D_\tau \theta^\alpha + w_\alpha D_\tau \lambda^\alpha \Bigg\}.
\label{34}
\end{eqnarray}
It is also worthwhile to notice that the BRST transformation 
is nilpotent up to the 'gauge' transformations 
\begin{eqnarray}
\delta_G d_\alpha &=& \frac{1}{4} \left[ (D_\tau \theta \Gamma_c \lambda
+ \lambda \Gamma_c D_\tau \theta) (\Gamma^c \lambda)_\alpha
+  (\Gamma^c \lambda)_\alpha  (D_\tau \theta \Gamma_c \lambda
+ \lambda \Gamma_c D_\tau \theta)  \right] , \nn\\
\delta_G w_\alpha &=&  - \frac{i}{2} \left[ P_c (\Gamma^c\lambda)_\alpha
+ (\Gamma^c \lambda)_\alpha P_c \right],
\label{35}
\end{eqnarray}
which are indeed symmetry of the matrix model (\ref{34}). 
Let us notice that this matrix model is nothing but the matrix model 
which can be obtained from the pure spinor formalism of the 11D
superparticle \cite{Brink} by generalizing all the local fields to matrices.

We shall finally make comments on some features of matrix 
model (\ref{34}).
First of all, this matrix model is not only invariant under the space-time
supersymmetry and the Berkovits' BRST transformation but also manifestly
covariant under $SO(1,10)$ Lorentz group, which is the most appealing
point of the model at hand.  However, the matrix model does not have 
the potential term given by $([x^a, x^b])^2$ (which exists in 
$L_{\det \pi^2}$) as in the BFSS M(atrix) model so the physical
properties of the both models are quite different as shown in the
next section.

\rm
\section{The one loop effective potential}

In this section, we wish to clarify the physical properties of our new
matrix model of 11D superparticle. It is well known that the
superparticle action in the continuum theory \cite{Brink} is the 
zero-slope limit of the superstring theory \cite{GS}, so it might
hopefully shed some light on the underlying structure of space-time.
However, as shown below by evaluating the one-loop effective potential
the matrix theory of superparticle in the pure spinor formalism
is a free theory, so scattering amplitudes should be calculated
by determining the vertex operators and inserting them in the path
integral.

In order to evaluate the one-loop effective potential, we 
take the gauge condition $A_\tau = 0$ and introduce the FP ghosts
$(\bar{C}, C)$ \footnote{See \cite{Douglas, Becker, Okawa} for
calculations of the effective action in M(atrix) theory.}. 
After integrating over $A_\tau$, we obtain the
gauge-fixed, BRST-invariant action
\begin{eqnarray}
S &=& \int d \tau Tr  \Bigg\{ \frac{1}{2} \left(\partial_\tau x^c + \frac{i}{4} 
(\theta \Gamma^c \partial_\tau \theta - \partial_\tau \theta \Gamma^c
\theta) \right)^2 + d_\alpha \partial_\tau \theta^\alpha 
+ w_\alpha \partial_\tau \lambda^\alpha - \bar{C} \partial_\tau C \Bigg\}.
\label{36}
\end{eqnarray}
As a background, we select a non-trivial classical solution 
\begin{eqnarray}
x^1_{(0)} = \frac{1}{2} \left(
\begin{array}{cc}
v \tau & 0 \\
0  & - v \tau
\end{array}
\right), \
x^2_{(0)} = \frac{1}{2} \left(
\begin{array}{cc}
b & 0 \\
0  & - b
\end{array}
\right),
\label{37}
\end{eqnarray}
which describes two particles moving with velocities $v/2$ and $- v/2$
and separated by the distance $b$ along the $x^2$-th axis. Around this
background, we expand $x^c$ by $x^c = x^c_{(0)} + y^c$ where
the fluctuation $y^c$ takes the off-diagonal form
\begin{eqnarray}
y^c = \left(
\begin{array}{cc}
0 & y^c \\
y^{\dagger c}  & 0
\end{array}
\right).
\label{38}
\end{eqnarray}
Similarly, $C$, $\bar{C}$, $w_\alpha$, $\lambda^\alpha$, $p_\alpha$
and $\theta^\alpha$ are expanded in the off-diagonal form like $y^c$.
(For convenience, we have used the same letters as the original fields
for expressing the off-diagonal matrix elements.) 

After inserting these equations into the action (\ref{34}) and
taking the quadratic terms with respect to the fluctuations,
we obtain the following action:
\begin{eqnarray}
S_2 &=& \int d \tau Tr \ \Bigl(  - y^\dagger_c \partial^2_\tau y^c
+ p_\alpha \partial_\tau \theta^{\dagger\alpha}
+ p^\dagger_\alpha \partial_\tau \theta^\alpha
+ w_\alpha \partial_\tau \lambda^{\dagger\alpha}
+ w^\dagger_\alpha \partial_\tau \lambda^\alpha \nn\\
&-& \bar{C} \partial_\tau C^\dagger - \bar{C}^\dagger \partial_\tau C
\Bigr).
\label{39}
\end{eqnarray}
In deriving this quadratic action, we have used the fact that
in the one-loop approximation, we can put $P^c = \partial_\tau
x^c_{(0)}$. Then the partition function is given by
\begin{eqnarray}
Z &=& \int {\cal D}X \ \e^{- S_2} \nn\\
&=& (\det \partial_\tau^2)^{-11} (\det \partial_\tau)^{-46}
(\det \partial_\tau)^{64} (\det \partial_\tau)^2 (\det \partial_\tau)^2 \nn\\
&=& (\det \partial_\tau)^{-22-46+64+2+2} \nn\\
&=& 1,
\label{40}
\end{eqnarray}
where we have symbolically denoted the integration measure by ${\cal
D}X$ and taken account of the contribution from the missing ghosts 
$(b, c)$ \cite{Ber8} in the pure spinor formalism. The result shows 
that at least in the one-loop level the theory is trivial, 
in other words, two particles do not interact with each other
\footnote{One subtle point of the above calculation is that we
have taken the 'axial' gauge $A_\tau = 0$. It is known that the effective
potential in general depends on the gauge conditions whereas the
S-matrix does not depend on the gauge. To have the gauge-invariant
effective action, we usually take the background field-dependent gauges
as in \cite{Douglas, Becker, Okawa}, which guarantees the gauge
invariance at all the stage of calculations. Our result obtained above,
however, is manifestly gauge-invariant so it is free from the problem of
the gauge dependence.}. Recall that in M(atrix) theory by
BFSS the similar calculation leads to the phase shift of D-particles in
the eikonal approximation  \cite{BFSS, Douglas, Becker, Okawa}. 
Our matrix theory therefore seems to be a free theory owing
to the lack of the potential term $([x^a, x^b])^2$. Thus, in order
to have non-trivial physical scattering amplitudes we must
evaluate the expectation values of the vertex operators even in the
matrix theory.
 
Finally, let us ask ourselves why we have obtained the matrix
model (\ref{34}) which is quite different from the BFSS matrix model.
First, we should notice that the transformation law of the supersymmetry
is completely different in both the formalisms. That is, our law
(\ref{30}) is purely from supermembrane whereas their law is from
super Yang-Mills theory \cite{BFSS}
\begin{eqnarray}
\delta X^i &=&  - 2 \epsilon^T \gamma^i \theta,  \nn\\
\delta \theta &=& \frac{1}{2} \left( D_t X^i \gamma_i + \gamma_{-}
+ \frac{1}{2} [X^i, X^j] \gamma_{ij} \right) \epsilon + \epsilon' \nn\\
\delta A_0 &=& -2 \epsilon^T \theta,
\label{41}
\end{eqnarray}
where $\epsilon$ and $\epsilon'$ are two independent 16 component
constant parameters. (Note that $A_0$ is needed to make the algebra of
supersymmetry close.) 
Second, the Berkovits' BRST invariance plays a role similar to a local
symmetry in the matrix model, thereby strongly restricting the form
of the action of the matrix model. In particular, the non-trivial
potential $([x^a, x^b])^2$ in M(atrix) theory, which is also
present in the $L_{\det \pi^2}$ in Eq. (\ref{31}), is not allowed to
satisfy the matrix version of the Berkovits' BRST symmetry.
In any case, since the symmetries in both the present matrix model 
and the BFSS M(atrix) theory are different so that the two theories belong
to different universality classes, it is natural to obtain the different
theories in the both approaches.

\rm
\section{Conclusion}

In this article, we have investigated the possibility of making use of 
the Berkovits pure spinor formalism in order to make a Lorentz covariant
M(atrix) theory. We have clarified that the naive expectation of it
does not work well since symmetries in the Berkovits pure spinor
formalism and the BFSS M(atrix) theory are different.
Moreover, we have pointed out that 
the Berkovits' BRST symmetry excludes the presence of the potential
$([x^a, x^b])^2$ which not only leads to an interesting interpretation
of space-time relevant to the non-commutative geometry but also
produces the non-trivial interaction of 11D supergravitons in M(atrix)
theory. Instead, we have constructed a matrix model of 11D
superparticle which is in a sense a particle limit of M(atrix)
theory.

Obviously we have many remaining future works to be investigated.
For instance, we have not constructed the vertex operators which
should be also invariant under the Berkovits' BRST transformation.
Related to this work, there is a computation of scattering
amplitude using superparticle in the continuum theory \cite{Green}
where the amplitude leads to a divergent result and the coefficient
is fixed by using the duality of superstring theory. We think that
once the vertex operators are constructed in the present formalism,
the scattering amplitude can be calculated and gives rise to
a finite result. This study is under investigation
and we wish to report the results in future publication.

\vs 1

\begin{flushleft}
{\bf Acknowledgements}
\end{flushleft}

We are grateful to N. Berkovits and M. Tonin for valuable discussions 
and would like to thank Dipartimento di Fisica, Universita degli 
Studi di Padova for its kind hospitality.
This work has been partially supported by the grant from 
the Japan Society for the Promotion
of Science, No. 14540277.

\vs 1

\end{document}